\documentclass[showpacs,amsmath,amssymb,pre,aps,superscriptaddress,twocolumn]{revtex4-1}
\usepackage[tight,TABTOPCAP]{subfigure}
\usepackage[usenames, dvipsnames]{color}
\usepackage{natbib}
\usepackage[utf8]{inputenc}
\usepackage{textcomp}
\usepackage[english]{babel}
\usepackage{amsmath}
\usepackage{amsfonts}
\usepackage{amssymb}
\usepackage{makeidx}
\usepackage{graphicx}
\usepackage{tikz}
\usepackage{float}
\usetikzlibrary{arrows}
\usetikzlibrary{decorations.pathreplacing,decorations.markings}
\usetikzlibrary{mindmap, trees}
\usetikzlibrary{shapes.geometric}
\usepackage{pgfplots}
\pgfplotsset{compat=1.16}
\usetikzlibrary{3d}
\usepackage[left=2cm,right=2cm,top=2cm,bottom=2cm]{geometry}
\usepackage{hyperref}

\begin{document}

\title{Connecting density fluctuations and Kirkwood-Buff integrals for finite-size systems}
\author{Mauricio Sevilla}
\affiliation{Max Planck Institute for Polymer Research, Ackermannweg 10, 55128, Mainz}

\author{Robinson Cortes-Huerto}
\email{corteshu@mpip-mainz.mpg.de}
\affiliation{Max Planck Institute for Polymer Research, Ackermannweg 10, 55128, Mainz}
\date{\today}
\begin{abstract}
Kirkwood-Buff integrals (KBI) connect the microscopic structure and thermodynamic properties of liquid solutions. KBI are defined in the grand canonical ensemble and evaluated assuming the thermodynamic limit (TL). In order to reconcile analytical and numerical approaches, finite-size KBI have been proposed in the literature, resulting in two strategies to obtain their TL values from computer simulations. (i) The spatial block-analysis method in which the simulation box is divided into subdomains of volume $V$ to compute fluctuations of the number of particles. 
(ii) A direct integration method where a corrected radial distribution function and a kernel that accounts for the geometry of the integration subvolumes are combined to obtain KBI as a function of $V$. 
In this work, we propose a method that connects both strategies into a single framework. We start from the definition of finite-size KBI, including the integration subdomain and an asymptotic correction to the radial distribution function, and solve them in Fourier space where periodic boundary conditions are trivially introduced. The limit $q\to 0$, equivalent to the value of the KBI in the TL, is obtained via the spatial block-analysis method.  When compared to the latter, our approach gives nearly identical results for all values of $V$. Moreover, all finite-size effect contributions (ensemble, finite-integration domains and periodic boundary conditions) are easily identifiable in the calculation. This feature allows us to analyse finite-size effects independently and extrapolate the results of a single simulation to different box sizes. To validate our approach, we investigate prototypical systems, including SPC/E water and aqueous urea mixtures.    	
\end{abstract}
\maketitle
\section{Introduction}
 
Kirkwood-Buff integrals (KBI) connect the microscopic structure of a liquid solution, via integrals of the radial distribution functions (RDF), and its thermodynamic properties, as obtained from fluctuations of the number of particles  in subvolumes of the total system \cite{KirkwoodBuff1951}.  This connection between local structure and thermodynamics is particularly useful in computational soft-matter studies where KBI are widely used to evaluate isothermal compressibilities, partial molar volumes and derivatives of chemical potentials \cite{ben-naim, CortesHuerto_Entropy2018, Robin2018b,Braten_JCIM61_840_2021}.  In particular, applications of KBI include the investigation of the thermodynamics of complex molecular mixtures~\cite{Mukherji_etal_JCTC8_375_2012,Mukherji_etal_JCTC8_3536_2012,GALATA201825,Narayanan_etal_JPCC122_10293_2018,Petris_etal_JPCB123_247_2019,LOVRINCEVIC2019111447,Celebi_etal_JCP154_184502_2021}, solvation of macromolecules \cite{KANG200714,Pierce2008,SHIMIZU2017128,Mukherji_Kremer_Macromol46_9158_2013,Oprzeska-Zingreb_Smiatek_BiophysRev10_809_2018,Oprzeska-Zingrebe_etal_EPJST227_1665_2019,nano10081460,D0CP05356B}, multicomponent diffusion in liquids \cite{Kjelstrup-etal-2014,Liu-etal-2013}, protein self-assembly and aggregration \cite{ProSA,Gee_Smith_JCP131_165101_2009}, Hofmeister ion chemistry~\cite{Bruce_etal_JACS142_19094_2020}, identification of nanostructures in water solutions of ionic liquids~\cite{Kumari_etal_PCCP23_944_2021} and the parameterisation of atomistic~\cite{Gee_etal_JCTC7_1369_2011,Fyta_EPJE35_21_2012,Schneck_etal_JPCB117_8310_2013,Loche_etal_JPCB125_8581_2021} and coarse-grained~\cite{Ganguly_etal_JCTC8_1802_2012,deOliveira_etal_JCP144_174106_2016} force fields.  Recently, KBI have been applied to compute isothermal compressibilities of prototypical crystals~\cite{Krueger_PRE103_L061301_2021,Miyaji_etal_JCP154_164506_2021}, showing unprecedented flexibility and range of applicability. 

KBI are strictly defined in the grand canonical ensemble. Moreover, in practice, it is usual to take the thermodynamic limit (TL) to reduce their calculation to spherically symmetric real-space integrals of the radial distribution functions. In computer simulations, the TL is approximated by introducing periodic boundary conditions (PBC) for a system with a fixed number of particles $N_0$. Accordingly, finite subvolumes $V$ with an average number of particles $\langle N \rangle$ are used to compute fluctuations of the number of particles and radial distribution functions~\cite{hill,Site_2017}. Periodicity, different thermodynamic ensembles and finite integration domains introduce artefacts in the resulting KBI.

Finite-size KBI have been proposed in the literature to bridge the existing gap between analytical expressions and numerical studies. The critical assumption is that fluctuations of the number of particles in subvolumes $V$ inside the simulation box are equal to integrals of the corresponding closed-system radial distribution functions~\cite{Binder-etal-EPL6-585-1988,Binder-etal-JPhysCondensMatter2-7009-1990,Roman-etal-JChemPhys107-4635-1997}. This equality provides two routes to obtain KBI in the TL. The first one, i.e. the spatial block analysis method (SBA), is based on calculating fluctuations of the number of particles in subdomains of volume $V$.  By using arguments from thermodynamics of small systems~\cite{hill}, linear scaling relations are defined to extrapolate KBI in the TL~\cite{SchnellChemPhysLett504-199-2011,Vlugt2}. The second possibility is to correct the radial distribution functions for the differences in the thermodynamic ensemble, then integrate them using a kernel that takes into account finite-size domains~\cite{schnell-etal-JPhysChemLett4-235-2013,ganguly-vandervegt-JCTC9-1347-2013}. Naturally, the limit $V\to \infty$ gives the KBI in the TL. 

Indeed, the two approaches are connected. In the limit $V>V_{\zeta}$, with $V_{\zeta}$ the volume defined by the correlation length of the system $\zeta$, the integration of the radial distribution functions give an expression equivalent to the result obtained from linear scaling of fluctuations of the number of particles, including ensemble and finite integration domain effects~\cite{CortesHuerto_Communication2016}.  Nevertheless, the local solvation structure information, provided by the short-range part of the RDF, is lost in this case.  Moreover, the effect of periodic boundary conditions is not included in the final result. 

In this work,  we propose a method that connects the spatial block analysis method to the direct integration of exact, finite-size KBI. We evaluate the large $r$ limit by introducing an asymptotic correction to the RDF.  By defining the geometry of the subdomain, we write and solve KBI in Fourier space where the periodicity of the cell can also be incorporated,  following the procedure proposed in Ref.~\onlinecite{Roman1999}. We compute the $q\to 0$ limit by using the spatial block analysis method. We thus obtain KBI as a function of the volume of the subdomain and find excellent agreement with fluctuations of the number of particles for SPC/E water and aqueous urea mixtures for all values of $V$.  The method is accurate, and its implementation is straightforward. It simply requires performing a spatial block analysis and calculating one-dimensional integrals of partial structure factors.

The paper is organised as follows:  In Section  \ref{Sec:M} we introduce the method, and in Section \ref{Sec:C} the computational details. We present the main results in Section \ref{Sec:R} and conclude in Section \ref{Sec:Con}.

\section{Kirkwood-Buff integrals for finite-size systems}\label{Sec:M}

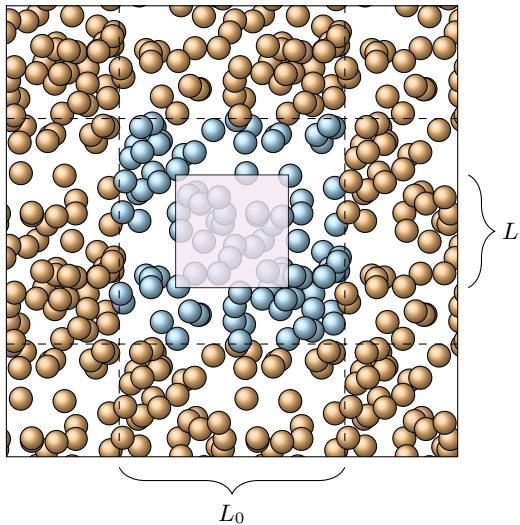
\begin{figure}[ht]

\definecolor{myblue}{RGB}{0,162,255}
\definecolor{myorange}{RGB}{255,137,0}
\centering
\begin{tikzpicture}

\begin{scope}[scale=0.75]

\begin{scope}
\clip (-4,-4) rectangle (4,4.);{

\foreach \i in {1,...,101}
{
\coordinate (A) at (2.*rand,2.*rand);
\node [draw,ball color=myblue!30,circle] at (A) {};

\node [draw,ball color=myorange!50,circle,shift={(3,0)}] at (A) {};
\node [draw,ball color=myorange!50,circle,shift={(-3,0)}] at (A) {};
\node [draw,ball color=myorange!50,circle,shift={(3,3)}] at (A) {};
\node [draw,ball color=myorange!50,circle,shift={(-3,3)}] at (A) {};
\node [draw,ball color=myorange!50,circle,shift={(0,3)}] at (A) {};
\node [draw,ball color=myorange!50,circle,shift={(0,-3)}] at (A) {};
\node [draw,ball color=myorange!50,circle,shift={(3,-3)}] at (A) {};
\node [draw,ball color=myorange!50,circle,shift={(-3,-3)}] at (A) {};
}}

\end{scope}
\draw [dashed] (-2,-4) -- (-2,4)  (2,-4) -- (2,4)  (-4,-2) -- (4,-2) (-4,2) -- (4,2);
\draw  (-4,-4) rectangle (4,4.);
\draw [decorate,decoration={brace,amplitude=10pt,mirror,raise=4pt},yshift=0pt]
(-2,-4.) -- (2,-4.) node [above,midway,yshift=-1.cm] {$L_0$};
draw [fill=violet!20,opacity=0.8]  (-1,-1) -- (1,-1) -- (1,1)  -- (-1,1) -- cycle;
\draw [decorate,decoration={brace,amplitude=10pt,mirror,raise=4pt},yshift=0pt]
(4.,-1.) -- (4.,1.) node [midway,xshift=.7cm] {$L$};

\draw [fill=violet!10,opacity=0.8]  (-1,-1) -- (1,-1) -- (1,1)  -- (-1,1) -- cycle;

\end{scope}

\end{tikzpicture}
\caption{Schematic representation of the spatial block analysis method. The $N_0$  blue particles represent the system with linear size $L_0$, and the red particles represent the periodic images. The purple box is a subvolume of linear size $L< L_0$ defined to compute fluctuations of the number of particles.}\label{fig:SBA}
\end{figure}

For a multicomponent fluid of species $i, j$, contained in a volume $V=L^3$, in thermal and chemical equilibrium with an infinite reservoir of particles, the Kirkwood-Buff integral (KBI) is defined as~\cite{KirkwoodBuff1951}
\begin{equation}\label{eq:KBI}
\begin{aligned}
G_{i j} &=V\left(\frac{\left\langle N_{i} N_{j}\right\rangle-\left\langle N_{i}\right\rangle\left\langle N_{j}\right\rangle}{\left\langle N_{i}\right\rangle\left\langle N_{j}\right\rangle}-\frac{\delta_{i j}}{\left\langle N_{i}\right\rangle}\right) \\
&=\frac{1}{V} \int_{V} \int_{V} d \mathbf{r}_{1} d \mathbf{r}_{2}\left[g_{i j}\left(\mathbf{r}\right)-1\right]\, ,
\end{aligned}
\end{equation}
where $N_i$ is the number of particles of the $i$-species and the bracket $\langle \cdots \rangle$ denotes a thermal average, $\delta_{ij}$ is the Kronecker delta and $g_{ij}$ is the pair correlation function defined in the grand canonical ensemble with $\mathbf{r}=\mathbf{r}_2-\mathbf{r}_1$.\\
In computer simulations, we usually investigate systems with fixed number of particles $N_0$ with volume $V_0=L^3_0$. Building on similar results for the Ornstein-Zernike equation~\cite{Roman1999}, we define the finite-size KBI as
\begin{equation}
\begin{split}
\label{eq:KBI_closed}
G_{ij}(V;V_{0}) &= V\left ( \frac{\langle N_{i}N_{j}\rangle' - 
\langle N_{i}\rangle' \langle N_{j}\rangle'}{\langle N_{i}\rangle' \langle N_{j}\rangle'} - 
\frac{\delta_{ij}}{\langle N_{i}\rangle' } \right )\\
= \frac{1}{V}&\int\, \int\, d\mathbf{r}_{1}
d\mathbf{r}_{2}\, R(\mathbf{r}_1)\, R(\mathbf{r}_2)\, [g_{ij}( \mathbf{r};V_{0} ) - 1]\, ,
\end{split}
\end{equation}
where the average $\langle \cdots \rangle' \equiv \langle \cdots \rangle_{V,V_{0}}$ now explicitly depends on the subdomain and total volumes, $V$ and $V_0$, respectively (See Fig.~\ref{fig:SBA}). Here we focus on the integral term, that contains the radial distribution function of the closed system, $g_{ij}(\mathbf{r};V_0)$, and a step function $R(\mathbf{r})$ that defines the integration subdomain: it is one inside and zero outside the volume $V$.  By defining  Eq.~\eqref{eq:KBI_closed}, we connect explicitly density fluctuations  and the integral of the pair correlation function for any subdomain $V$. \\
In the following, we focus on integrating the r.h.s. of Eq.\ \eqref{eq:KBI_closed}. That is
\begin{equation}
G_{ij}(V;V_{0}) =  \frac{1}{V}\int\, \int\, d\mathbf{r}_{1}
d\mathbf{r}_{2}\, R(\mathbf{r}_1)\, R(\mathbf{r}_2)\, h_{ij}( \mathbf{r};V_{0} )\, ,
\end{equation}
with $h_{ij}( \mathbf{r};V_{0} ) = g_{ij}( \mathbf{r};V_{0} ) - 1$.\\
To include the correction due to ensemble effects, we use the approximation proposed in Ref.~\onlinecite{CortesHuerto_Communication2016}
\begin{equation}
\label{eq:corr_rdf}
g_{ij}(\mathbf{r};V_{0}) = g_{ij}(\mathbf{r}) 
-\frac{1}{V_{0}}\left(\frac{\delta_{ij}}{\rho_{i}} + G_{ij}^{\infty}\right)\, ,
\end{equation}
based on the asymptotic limit 
$g_{ij}(r\to \infty;V_{0}) = 1 -
(\delta_{ij}/\rho_{i} + G_{ij}^{\infty})/V_{0}$ discussed in Ref.~\onlinecite{ben-naim}. This implies that 
\begin{equation}
 h_{ij}( \mathbf{r};V_{0} ) = h_{ij}(\mathbf{r})  -\frac{1}{V_{0}}\left(\frac{\delta_{ij}}{\rho_{i}} + G_{ij}^{\infty}\right)\, ,  
\end{equation}
thus, the finite-size KBI becomes
\begin{equation}\label{eq:KBI_All}
  G_{ij}(V;V_{0}) =  G_{ij}(V) - \frac{V}{V_{0}}\left(\frac{\delta_{ij}}{\rho_{i}} + G_{ij}^{\infty}\right) \, ,  
\end{equation}
where the second term on the r.h.s. contains the correction due to ensemble effects~\cite{Roman2008,CortesHuerto_Communication2016,CortesHuerto_Entropy2018,Robin2018b} and
\begin{equation}\label{eq:KBI_real}
  G_{ij}(V) =  \frac{1}{V}\int\, \int\, d\mathbf{r}_{1}
d\mathbf{r}_{2}\, R(\mathbf{r}_1)\, R(\mathbf{r}_2)\, h_{ij}( \mathbf{r}) \, .  
\end{equation}
This expression can be easily written in Fourier space
\begin{equation}\label{eq:KBI_Fourier}
G_{i j}(V) =\frac{1}{(2\pi)^3V} \int d \mathbf{k}\, \tilde{R}(\mathbf{k})\, \tilde{R}(-\mathbf{k})\,\tilde{h}_{i j}\left(\mathbf{k}\right)\, ,
\end{equation}
where $\tilde{h}_{i j}$ is the Fourier transform of $h_{ij}$. An additional advantage of integrating in reciprocal space is that periodic boundary conditions can be considered explicitly~\cite{Roman1999,Villamaina-Trizac-EurJPhys35-035011-2014}. It is enough to rewrite $\tilde{h}_{ij}(\mathbf{k})$ such that periodic copies of the system are included via a phase factor. That is, we include the complete contribution of the periodic boundary conditions into Eq.\ \eqref{eq:KBI_Fourier} as
\begin{equation}\label{eq:KBI_Fourier_PBC}
G_{i j}(V) =\frac{1}{(2\pi)^3V} \int d \mathbf{k}\, \tilde{R}(\mathbf{k})\, \tilde{R}(-\mathbf{k})\, \tilde{h}^{\rm PBC}_{ij}(\mathbf{k})\, ,
\end{equation}
where~\cite{Roman1999}
\begin{equation}\label{eq:hPBC}
\tilde{h}^{\rm PBC}_{ij}(\mathbf{k})=\sum_{n_x,n_y,n_z} e^{-\mathbf{k}\cdot\mathbf{s}_{n_x,n_y,n_z}} \tilde{h}_{ij}(\mathbf{k})\, ,
\end{equation}
with $\mathbf{s}_{n_x,n_y,n_z}=(n_x\, L_0,n_y\, L_0,n_z\, L_0)$ a vector specifying the system's periodic images such that $n_{x,y,z}$ takes integer values. In the following, we find that the choice $n_x=n_y=n_z=1$ is sufficient to compute Eq.~\eqref{eq:KBI_Fourier_PBC} accurately.\\

We assume a homogeneous fluid such that $\tilde{h}_{ij}(\mathbf{k})=\tilde{h}_{ij}(k)$ with $k=\sqrt{\mathbf{k} \cdot \mathbf{k} }$. Hence, in practice, we use the relation between $\tilde{h}_{ij}(k)$ and the partial structure factors $S_{ij}$~\cite{Ashcroft_Structure1967,Sij}, namely

\begin{equation}\label{eq:Sij}
    S_{ij}(k) = \delta_{ij} +\tilde{h}_{i j} (k)\, .
\end{equation}

The partial structure factors are computed as 

\begin{equation}
S_{ij}(\mathbf{k}) = \left\langle  \frac{1}{\sqrt{N_i N_j}}\sum_{i'=1}^{N_i}\sum_{j'=1}^{N_j}\exp{(-i\mathbf{k}\cdot(\mathbf{r}_{i'} - \mathbf{r}_{j'} ))} \right\rangle\, .
\end{equation}

Consequently, the problem reduces to evaluate a single integral of the partial structure factors given by Eq.~\eqref{eq:KBI_Fourier_PBC}.\\

In principle, Eq.\ \eqref{eq:KBI_All} now includes all the finite size effects present in the simulation (finite boundary, periodicity of the box and ensemble). Before entering into applications, there are still two issues demanding our immediate attention. The first is that the asymptotic correction in Eq.~\eqref{eq:KBI_All} requires the value of $G_{ij}^{\infty}$. The second concerns the evaluation of $\lim_{k\to 0} S_{ij}(k)$, that reduces, again, to evaluate $G_{ij}^{\infty}$. Indeed, we have

\begin{equation}
   \lim_{k\to 0} S_{ij}(k) = \delta_{ij} + \rho_i \, G_{ij}^{\infty}\, . 
\end{equation}

To obtain $G_{ij}^{\infty}$ we recall that, in the limit $V_{\zeta}< V< V_0$ (grand canonical ensemble), Eq.~\eqref{eq:KBI_real} can  be approximated to $G_{ij}(V)\approx G_{ij}^{\infty} + \alpha_{ij}/V^{1/3}$~\cite{Binder-etal-EPL6-585-1988,Binder-etal-JPhysCondensMatter2-7009-1990,SchnellChemPhysLett504-199-2011,schnell-etal-JPhysChemLett4-235-2013} where $\alpha_{ij}$ is a constant. By including this approximation into Eq.~\eqref{eq:KBI_All}, we recover the spatial block analysis (SBA) result consistent with the result reported in Ref.~\cite{CortesHuerto_Communication2016}
\begin{equation}
   G^{\rm SBA}_{ij}(V;V_{0}) =  G_{ij}^{\infty}\left( 1 - \frac{V}{V_{0}}\right) - \frac{V}{V_{0}}\frac{\delta_{ij}}{\rho_{i}} + \frac{\alpha_{ij}}{V^{1/3}}\, . 
   \label{eq:SBA_kirkwood}
\end{equation}
By evaluating density fluctuations for volumes $V\le V_0$,  as defined by the left hand side of Eq.~\eqref{eq:KBI_closed}, it is thus possible to extrapolate $G_{ij}^{\infty}$~\cite{SchnellChemPhysLett504-199-2011,CortesHuerto_Communication2016}.

To summarise, the present method to evaluate KBI for finite systems requires information readily accessible from the simulation trajectory: density fluctuations for subvolumes $V\le V_0$ and partial structure factors. Additional corrections to the RDF or finite-domain integration kernels are not required. Moreover, periodic boundary effects are trivially included in the calculation.

\section{Computational details}\label{Sec:C}
To validate our approach, we first focus on liquid SPC/E water~\cite{SPC1,*SPC2,*SPC3}. 
Molecular dynamics simulations have been carried out with GROMACS 4.5.1 \cite{GROMACS} for systems containing 1000 and 8000 water molecules. We started with systems of initial density $\approx$ 26 waters/nm$^3$ ($\approx$ 776\, kg/m$^3$) that were optimised using steepest descent minimisation (50000 steps are sufficient). An equilibration run of 3.5 ns was carried out in the NPT ensemble at 1 bar. Next, we alternated 3.5 ns (time step = 1 fs) 
 constant pressure (NPT) at P=1 bar and 
 constant volume (NVT) simulations at T=300 K. For NPT simulations we used the Berendsen barostat 
 \cite{Berendsen-etal-JChemPhys81-3684-1984}, and for 
 NVT simulations temperature was enforced by a velocity rescaling thermostat \cite{bussi2007canonical}. 
 We continued with this 
 protocol until we verified that in the NPT ensemble the density is 33.5 waters/nm$^3$ 
 (1000 kg/m$^3$)
 and that in the NVT simulation pressure fluctuates around the 1 bar 
 value. The last NVT trajectory obtained after this sequence of NPT--NVT equilibration runs was used for the spatial block analysis and for the calculation of the structure factor.

To test the method with a multicomponent case, we have re-used our simulation trajectories of aqueous urea solution \cite{deOliveira_etal_JCP144_174106_2016,Robin2018b} using the Kirkwood-Buff derived force field \cite{Weerasinghe2003} and SPC/E 
water \cite{SPC1,*SPC2,*SPC3} in GROMACS 4.5.1 \cite{GROMACS} with a relatively small size of the simulation box ($L\sim 8$ nm). We have 
considered four more molar concentrations for a total of seven molar concentrations: 2.00,  3.06,  3.90,  5.07, 6.03, 7.10 and 8.03 M. Hence, we have alternated 
3.5 ns (time step = 1 fs) 
 constant pressure (NPT) at P=1 bar and 
 constant volume (NVT) simulations at T=300 K. For NPT simulations we used a Berendsen barostat 
 \cite{Berendsen-etal-JChemPhys81-3684-1984} to control the pressure, and for 
 NVT simulations a velocity rescaling thermostat \cite{bussi2007canonical} to enforce the target temperature. 
 We continued with this 
 protocol (38 NPT--NVT cycles) until we verified that in the NVT simulation pressure fluctuates around 1 bar. Also in this case, the last NVT trajectory obtained after this NPT--NVT equilibration sequence was used for the spatial block analysis and for the calculation of the partial structure factors.

\section{Results}\label{Sec:R}
\subsection{Single component liquid: SPC/E water}

For the single-component liquid, we focus on the Ornstein-Zernicke integral equation for finite size systems~\cite{Binder1981,
Binder-etal-EPL6-585-1988,Salacuse-etal-PRE53-2382-1996,Roman-etal-EPL42-371-1998}. For a closed system with fixed number of particles $N_{0}$ and 
volume $V_{0}$, including PBC. Similar to Eq.~\ref{eq:KBI_closed}, we define \cite{Roman1999,Roman2008}

\begin{equation}
\label{eq:OZsize}
\begin{split}
&\chi_{T}(V;V_{0}) = \frac{\langle N^{2} \rangle' - 
\langle N \rangle'^{2} }{\langle N\rangle' }\\
&= 1 + \frac{\rho}{V}\int_{V}\int_{V}d\mathbf{r}_{1}
d\mathbf{r}_{2}R(\mathbf{r}_{1})R(\mathbf{r}_{2})[g(\mathbf{r};V_0) - 1]\, ,
\end{split}
\end{equation}

and in this case we use the asymptotic correction to the RDF proposed in Refs~\onlinecite{Percus1961II,Salacuse-etal-PRE53-2382-1996}

 \begin{equation}
  g(\mathbf{r};V_0) = g(\mathbf{r}) - \frac{\chi_{T}^{\infty}}{N_{0}}\, ,
 \end{equation}
by neglecting $O(1/N_{0}^{2})$ contributions. $\chi_{T}^{\infty} = \rho k_{\rm B}T \kappa_{T}$, $\kappa_{T}$ being the isothermal compressibility of the bulk system. To solve the integral on the r.h.s. of Eq.~\ref{eq:OZsize}, we use the same procedure as described in Section \ref{Sec:M}. In cases where $V_{\zeta}<V<V_0$, we obtain the equivalent spatial block analysis expression~\cite{CortesHuerto_Entropy2018,Robin2018b} 

\begin{equation}\label{eq:OZ_SBA}
 \chi_{T}^{\rm SBA}(V;V_0) = \chi_{T}^{\infty}\left( 1 - \frac{V}{V_0}\right) + \frac{\rho \alpha}{V^{1/3}}\, , 
\end{equation}

with $\alpha$ a constant. 

\begin{figure}[h]
\includegraphics[width=0.48\textwidth]{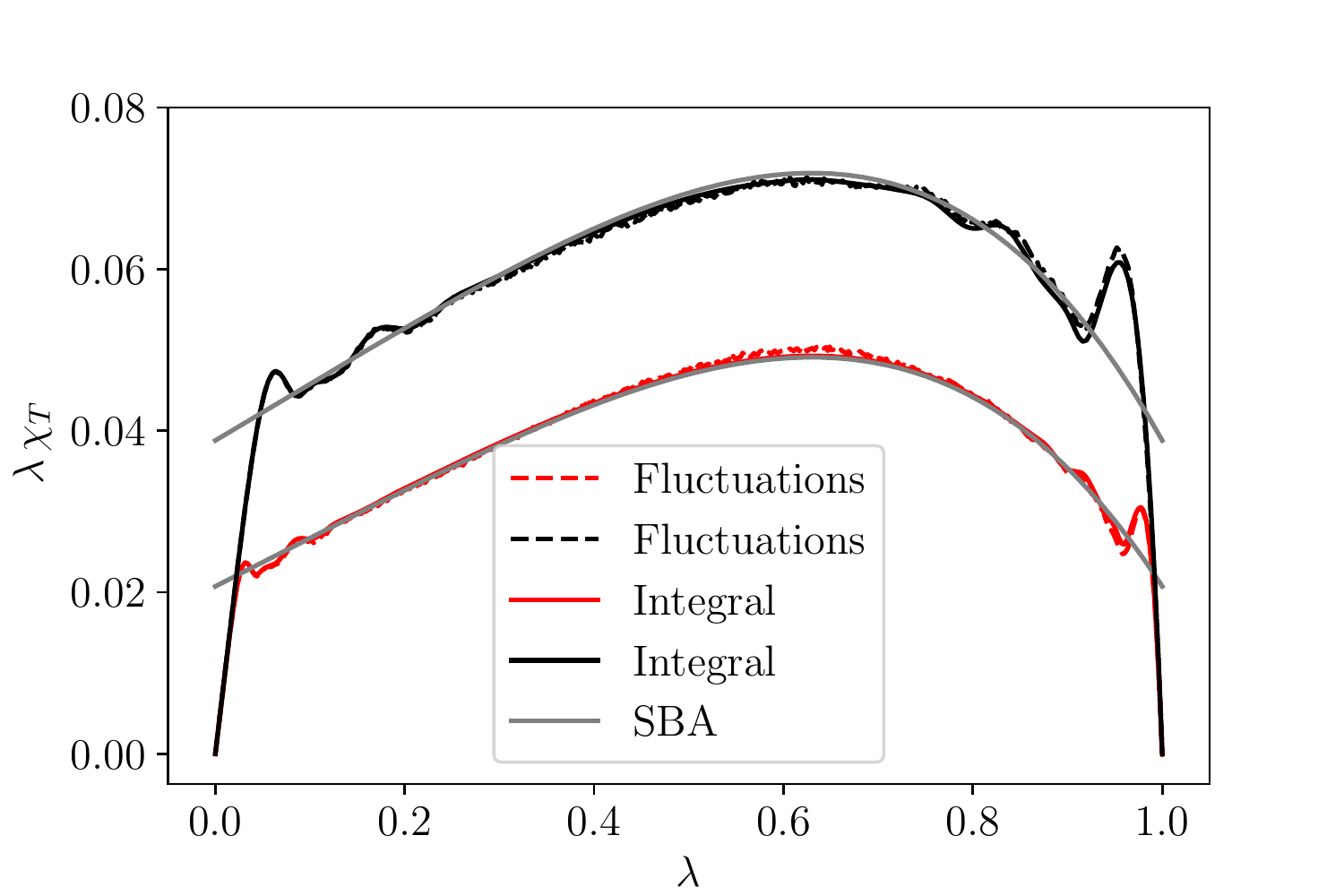}
\caption{Plots of the normalised finite-size isothermal compressibility $\lambda \chi_{T}$ as a function of $\lambda=(V/V_0)^{1/3}$ for systems  with $N_0=1000$ (black) and $8000$ (red) water molecules. Dashed lines correspond to density fluctuations and solid lines represent the method presented here. Solid grey lines correspond to the fitting of Eq.~\eqref{eq:OZ_SBA}}\label{fig:water_fluc_both}
\end{figure}

First, we compute density fluctuations as defined on the l.h.s. of  Eq.~\ref{eq:OZsize} for both $N_0=1000$ and $8000$ systems. By defining $\lambda=(V/V_0)^{1/3}$, we plot $\lambda \chi_{T}$ as a function of $\lambda$ (Fig.~\ref{fig:water_fluc_both}). We extrapolate $\chi_{T}^{\infty}$ from the curve's slope in the region $\lambda<0.3$ for the system with $N_0=8000$ water molecules. The choice of this linear region is motivated by the fact that $\lambda \chi_{T}$ as obtained from Eq.~\ref{eq:OZsize} has the maximum at $\lambda_{\rm max}=0.63$. Thus, we estimated $\lambda=0.3$ as the value where the curve starts deviating significantly from a straight line. We can also use $\lambda=0.3$ to choose an appropriate size of the system for the spatial block analysis. By assuming that the correlation length of water is $\zeta = 1.5$ nm, we define $V^{1/3}_{\zeta} = 1.5 \times (4\pi/3)^{1/3}$ nm. The simulation box with $N=8000$ water molecules has $V^{1/3}_0=6.2$ nm, thus $\lambda=(V_{\zeta}/V_0)^{1/3}=0.39$. This value is larger than $\lambda=0.3$, still, it is sufficient to obtain a value of $\chi_{T}^{\infty}=0.062$, in good agreement with the results reported in Ref.~\cite{Robin2018b}. In practice, to select the size of the system one can start by estimating the correlation length from the radial distribution function and evaluating the linear size of the box such that $\lambda \approx 0.3$. This criteria can also be applied to binary mixtures.

We use the results of this linear fit to plot the SBA results, Eq.~\eqref{eq:OZ_SBA}. Also for this system, we compute the structure factor, and correct  for  the  $\lim_{k\to 0}$ by  using the  relation 

\begin{equation}
    \lim_{k\to 0} S(k) = \chi_{T}^{\infty}\, .
\end{equation}

We thus compute an integral equivalent to Eq.~\eqref{eq:KBI_Fourier} to obtain $\chi_{T}(V;V_0)$. The results for both systems are also presented in Fig.~\ref{fig:water_fluc_both}. It is apparent that the agreement between density fluctuations and the integral method presented here is excellent. In contrast to the spatial block analysis result (Eq.~\eqref{eq:OZ_SBA}), oscillations of $\lambda \chi_{T}$ at low values of $\lambda$, related to the local liquid structure, and at large values of $\lambda$, due to the periodicity of the simulation box, are consistently reproduced with our method. This is particularly interesting for the system with $N_0=1000$ water molecules, where oscillations are more pronounced. In this case, our integral method uses information from the system with $N_0=8000$ water molecules. The small box behaviour is reproduced \emph{artificially} via the periodic images in Eq.~\eqref{eq:hPBC}.  

\begin{figure}[h]
\includegraphics[width=0.48\textwidth]{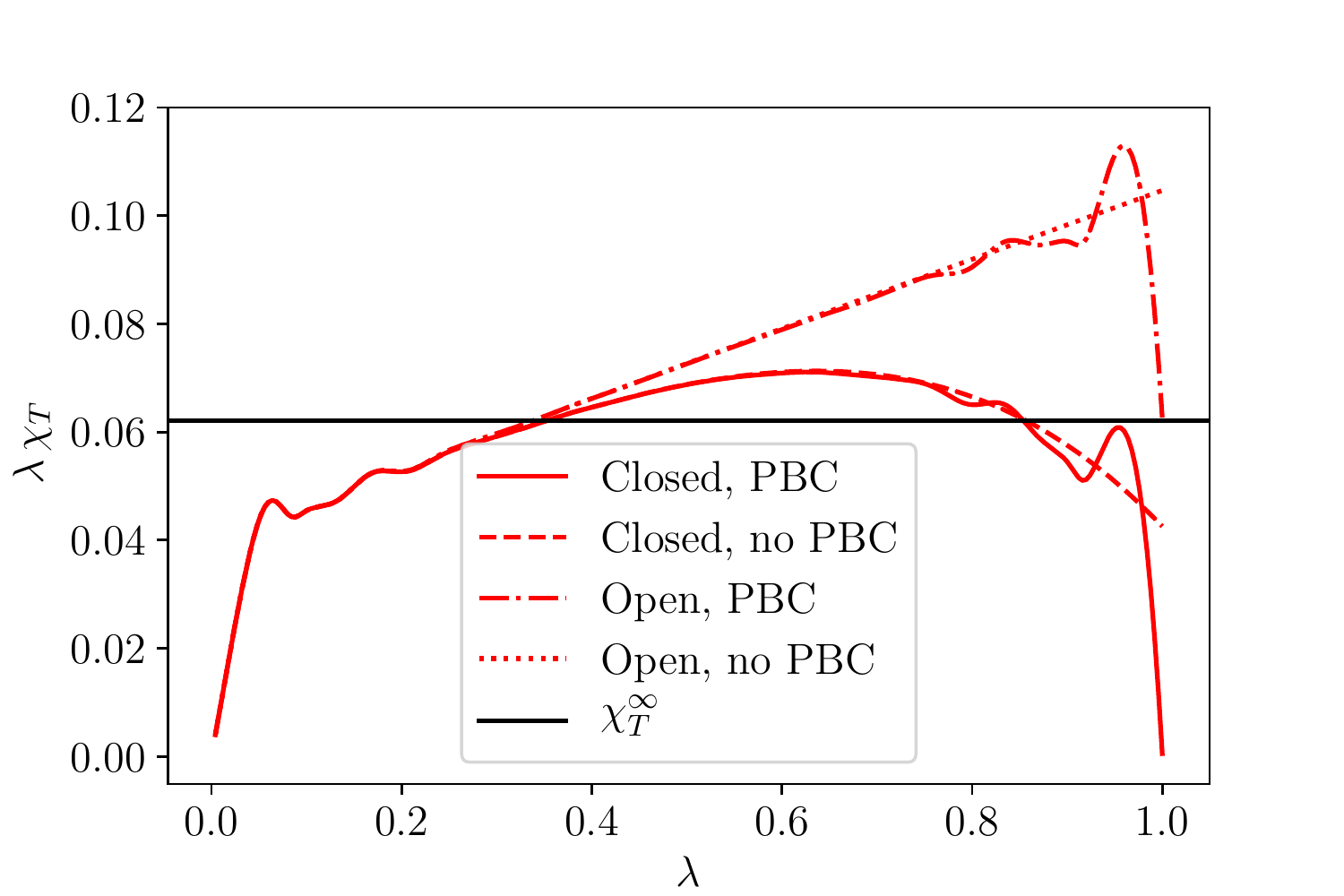}
\caption{Calculation of $\lambda \chi_{T}$ with our method, Eq. \ref{eq:KBI_Fourier_PBC}, for the system with $N_0=1000$ molecules. We present four cases: (solid) closed system with PBC, (dashed) closed system without PBC, (dash-dotted) open system with  PBC, (dotted) open system without PBC. The black horizontal line corresponds to $\chi_{T}^{\infty}=0.062$. }\label{fig:water_size_effects}
\end{figure}

We now focus on the different finite-size effects present in the system with $N_0=1000$ water molecules. In Fig.~\ref{fig:water_size_effects} we present four possibilities of evaluating the r.h.s. of Eq.~\eqref{eq:OZsize}. (i) For the closed system, i.e. including the correction $\chi_{T}^{\infty}V/V_0$, with PBC, we observe that $\chi_{T}(\lambda=1)=0$, as expected. (ii) The closed system without PBC gives a limit $\chi_{T}(\lambda=1)=\rho \alpha / V_0^{1/3}$ consistent with Eq.~\eqref{eq:OZ_SBA}. (iii) An open system can be obtained by neglecting the correction $\chi_{T}^{\infty}V/V_0$. Moreover, by including PBC, we obtain $\chi_{T}(\lambda=1)=\chi_{T}^{\infty}=0.062$, precisely the thermodynamic limit value. (iv) For an open system without PBC $\chi_{T}(\lambda=1)=\chi_{T}^{\infty}+\rho \alpha / V_0^{1/3}$, again, consistent with Eq.~\eqref{eq:OZ_SBA}. These results thus highlight the role of PBC in enforcing the correct behaviour at the boundary of open and closed molecular systems. 

\subsection{Binary mixture: aqueous urea solution}

\begin{figure*}[t!]
\includegraphics[width=0.8\textwidth]{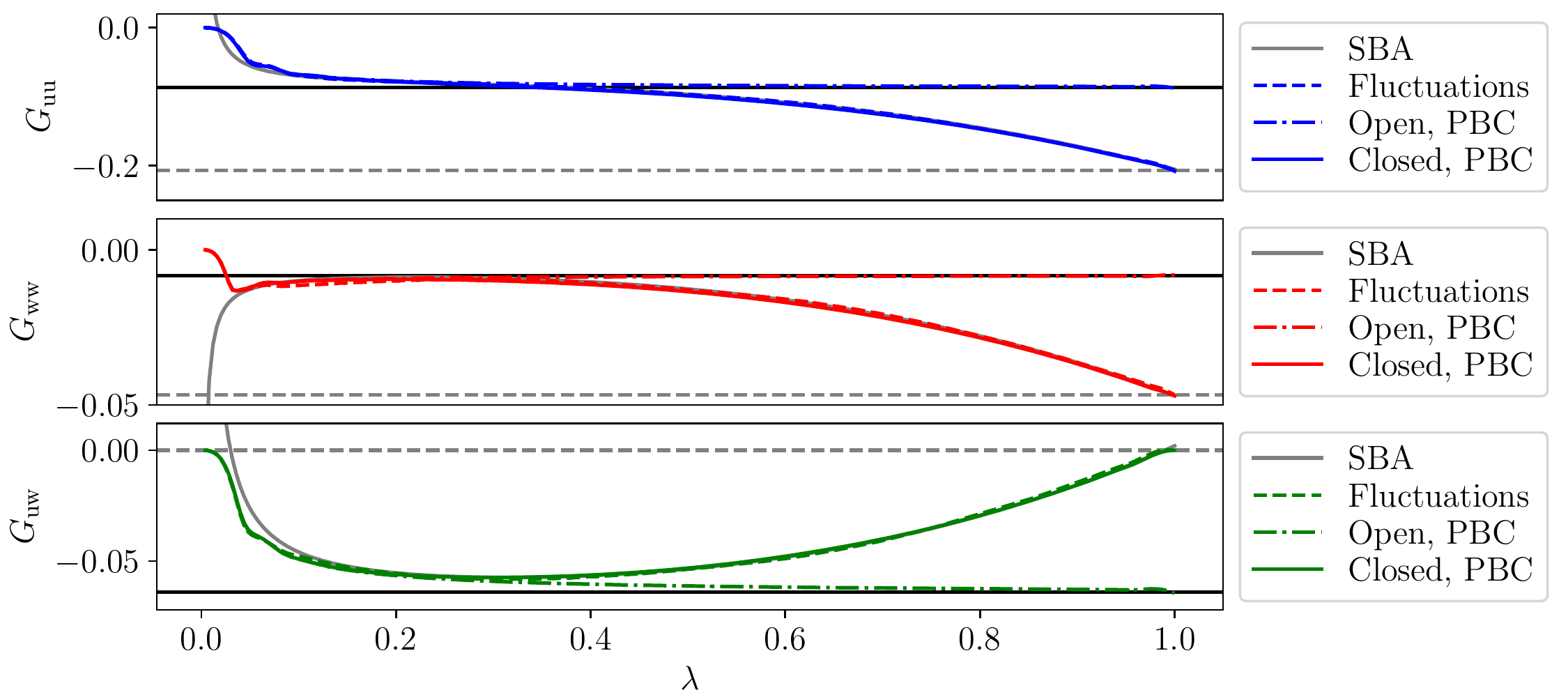}
\caption{KBI components $G_{\rm uu}$ (blue, top panel), $G_{\rm ww}$ (red, middle panel) and $G_{\rm uw}$ (green, bottom panel) for an 8M aqueous urea mixture as a function of $\lambda=(V/V_0)^3$. We present density fluctuations as obtained from the left hand side of Eq.~\eqref{eq:KBI_closed} (dashed lines), the spatial block analysis approximation in Eq.~\eqref{eq:SBA_kirkwood} (grey lines) and from the finite KBI expression, Eq.~\eqref{eq:KBI_All} with $G_{ij}(V)$ given by the Fourier integral Eq.~\eqref{eq:KBI_Fourier_PBC} with (solid) and without (dash-dotted lines) the correction to ensemble effects given by $G_{ij}^{\infty}\lambda^3$. The solid black lines correspond to the KBI in the TL, $G_{ij}^{\infty}$. The dashed grey lines indicate the asymptotic limit for the closed system, $-\delta_{ij}/\rho_i$. }\label{fig:mix_fluc}
\end{figure*}

We perform a similar analysis for the aqueous urea mixture case. First, we compute fluctuations of the number of particles as defined on the l.h.s. of Eq.~\eqref{eq:KBI_closed}.  As for the single component case, we define $\lambda=(V/V_0)^3$ and plot $\lambda G_{ij}$ as a function of $\lambda$. We carried out this study for all concentrations. However, we only present the results for the case 8M in Fig.~\ref{fig:mix_fluc} (dashed lines).  Using the information from the linear region $\lambda<0.3$, we extrapolate $G_{ij}^{\infty}$ and obtain $\alpha_{ij}$. In this case, we get $G_{\rm uu}^{\infty} = -0.0867$ ,$G_{\rm uw}^{\infty} = -0.0639$ and $G_{\rm ww}^{\infty} = -0.0083$ nm$^3$ with uu, uw and ww corresponding to urea-urea, urea-water and water-water components, respectively. These values well reproduce derivatives of activity coefficients reported experimentally~\cite{CortesHuerto_Communication2016,Robin2018b} as well as excess chemical potentials trends with concentration obtained with different computational methods~\cite{Kokubo2007,Baptista_DFT2021}. We insert these values in Eq.~\eqref{eq:SBA_kirkwood}, i.e. SBA, and plot this result as well (solid grey lines). Finally, we use the finite KBI introduced here (Eq.~\eqref{eq:KBI_All}) and use the Fourier integral, Eq.~\eqref{eq:KBI_Fourier_PBC} to compute $G_{ij}(V)$. We present both results with (solid lines) and without (dash-dotted lines) the correction to the ensemble finite-size effects, $G_{ij}^{\infty} \lambda^3$. 

In this case as well, the results of our method accurately reproduce density fluctuations in the whole range $0<\lambda<1$, including both, local structure ($\lambda\ll 1$) and periodic boundary ($\lambda \approx 1$) features. As anticipated, it is also apparent that the SBA result does not reproduce these limiting cases. Nevertheless, in the limit $\lambda =1$ the results from fluctuations, SBA and our integration (Closed, PBC) converge to $-\delta_{ij}/\rho_i$, the expected result of the KBI for a closed system~\cite{ben-naim}. As previously stated, we can separate finite-size contributions by focusing on the corresponding terms in Eqs~\eqref{eq:KBI_All} and \eqref{eq:KBI_Fourier_PBC}. In particular, for an open system (Open, PBC), i.e. $\lim_{V_0\to \infty}$, we verify that the KBI converge to $G_{ij}^{\infty}$ when $\lambda=1$.  

\begin{figure}[h]
\includegraphics[width=0.48\textwidth]{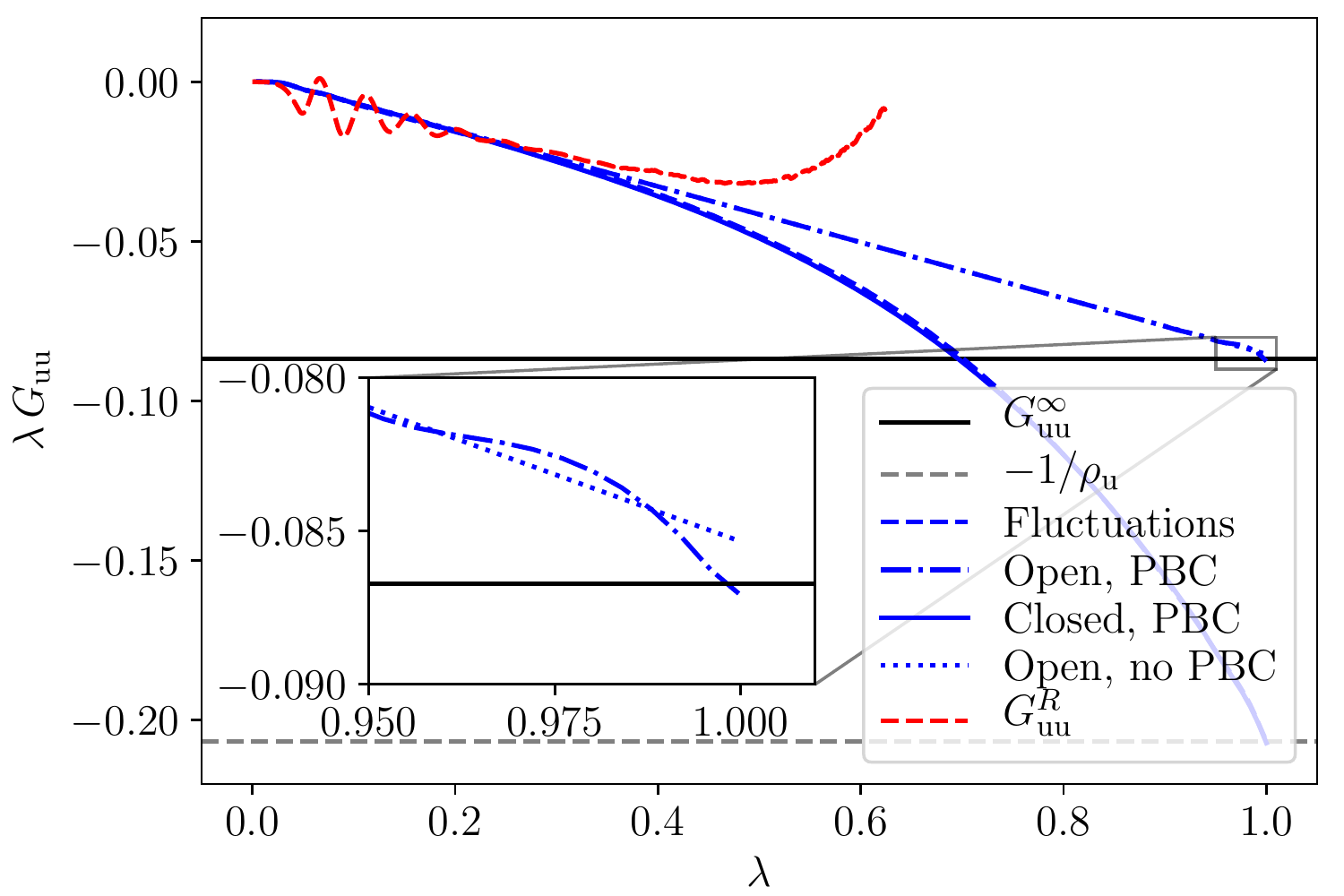}
\caption{Normalised KBI, $\lambda G_{\rm uu}$, as a function of $\lambda$ obtained using various methods and conditions. Fluctuations - Fluctuations of the number of particles as obtained from the l.h.s. of Eq.~\eqref{eq:KBI_closed} (dashed lines). Closed, PBC; Open, PBC - finite KBI expression, Eq.~\eqref{eq:KBI_All} with $G_{\rm uu}(V)$ given by the Fourier integral Eq.~\eqref{eq:KBI_Fourier_PBC} with (solid) and without (dash-dotted lines) the correction to ensemble effects given by $G_{\rm uu}^{\infty}\lambda^3$, both with PBC. Open, no PBC - finite KBI without the correction for the thermodynamic ensemble, and without PBC. The solid black line correspond to the KBI in the TL, $G_{\rm uu}^{\infty}$. The dashed grey line indicate the asymptotic limit for the closed system, $-1/\rho_{\rm u}$. The red dashed line corresponds to the running integral Eq.~\eqref{KBI_R}. (Inset) Detail of the convergence to the TL.}\label{fig:mix_limit}
\end{figure}

We examine this in more detail in Fig.~\ref{fig:mix_limit} where the normalised KBI for urea-urea, $\lambda G_{\rm uu}$, is presented. In addition to the limiting cases discussed above, we also consider an open system without periodic boundary conditions (dotted line). It is apparent in the region $\lambda\approx 1$ (inner panel) that $G_{\rm uu}$ for an open system with PBC converges to the value in the thermodynamic limit $G_{\rm uu}^{\infty}$, whereas for the open system without PBC, $G_{\rm uu}$ is slightly larger than $G_{\rm uu}^{\infty}$ value by a factor $\alpha_{\rm uu}/V_0^{1/3}$, as expected from the SBA expression, Eq.~\eqref{eq:SBA_kirkwood}. As in the single component case, this result emphasises that PBC enforce the correct behaviour at the boundary of closed and open liquid mixtures. Finally, we also present the normalised running integral $\lambda G_{ij}^R$ (dashed red line) using 

\begin{equation}\label{KBI_R}
    G_{ij}^R = 4\pi \int_{0}^R dr\, r^2 (g(r;V_{0})-1)\, ,
\end{equation}

(with $R>\zeta$), an expression frequently used in the literature. The major differences with the results presented in this work resulting from various finite-size effects highlight the apparent limitations of using such an expression to calculate KBI. 

\section{Concluding remarks}\label{Sec:Con}
Finite Kirkwood-Buff integrals (KBI) enable us to sample the thermodynamic limit of liquid mixtures via relatively small computer simulations. The definition of finite KBI balance fluctuations of the number of particles in subdomains within the simulation box and integrals of the corresponding RDF. In this work, we underline this equality by reproducing density fluctuations as a function of the linear size of the subdomain via a simple integration strategy. In particular, we introduce a method to evaluate KBI via integrals of the partial structure factors in reciprocal space. A significant advantage of our approach corresponds to the direct inclusion of finite integration domains and PBC contributions.
Consequently, we can now identify and \emph{remove} finite-size effects such that grand canonical and thermodynamic limit results become readily available from finite-size computer simulations. Moreover, we show that this scheme enables us to extrapolate our results to different sizes of the simulation box simply by modifying the periodicity factor in the integration procedure. The simplicity of the method is apparent since it only requires fluctuations of the number of particles calculated for different subdomains sizes and the partial structure factors. We foresee immediate applications in situations where PBC play a pivotal role, namely, the recently introduced KBI for crystalline materials~\cite{Krueger_PRE103_L061301_2021,Miyaji_etal_JCP154_164506_2021}. 

\acknowledgments
We are grateful to Kurt Kremer, Debashish Mukherji and Pietro Ballone for insightful discussions. We also thank Atreyee Banerjee for her critical reading of the manuscript. R.C.-H. gratefully acknowledges funding from SFB-TRR146 of the German Research Foundation (DFG). Simulations have been performed on the THINC cluster at the Max Planck Institute for Polymer Research and on the COBRA cluster of the Max Planck Computing and Data Facility. 
%
%
\bibliographystyle{unsrt}
\bibliography{bib.bib} 
\end{document}